# Analysis of Kak's Quantum Cryptography Protocol from the Perspective of Source Strength

Antony Akshay

**Abstract**
This paper analyzes the performance of Kak's quantum cryptography protocol when intensity monitoring is used to detect the presence of Eve during transmission. Some difficulties related to interception to obtain useful data from the transmission are discussed. The analysis shows the resilience of the protocol towards the man-in-the-middle attack.

**Introduction**
The field of quantum physics has opened up a new realm of cryptography. Quantum cryptography promises to offer perfect security if correctly implemented. The most prominent quantum cryptographic system is the BB84 [3] but now it is becoming clear that its implementation assumes unrealistic constraints which is why it has not turned out to be nearly as secure in practice [16],[17]. Basically, the problem with the implementation of BB84 is that it ideally requires that only one photon be transmitted for each key-bit and the detector be able to detect this single photon. Practical systems transmit a bunch of photon that create vulnerabilities that can be exploited by the eavesdropper.

In view of the difficulties of implementation of BB84, we consider Kak's quantum cryptography protocol (K06 protocol) [2] which is based on qubit (or polarization) rotations. Although qubit rotations [4]-[11] require careful control and knowledge of the initial state, the method still does not place the same constraints on sources and detectors as is required for BB84. We show that his protocol is resilient to the man-in-the-middle attack if intensity monitoring is use (which lends itself to this protocol in a natural fashion) and if the hash of the message to be sent is published.

**K06 Quantum Cryptography Protocol**
K06 was designed to transmit keys or data over the public channel between Alice (sender) and Bob (receiver.) Unlike BB84 that can only transmit keys, because of its slowness, the K06 protocol can be used to transmit whole messages [1]. This protocol involves the use of secret rotation transformations by both Alice and Bob that are not shared. Some considerations related to the protocol are given in the papers [12]-[14].

When Alice wants to send a message to Bob, she codes the information into the polarization of a photon (the state X) in a manner that is agreed upon in advance by Alice and Bob. Thus, for example, a horizontally polarized photon can be chosen to be the bit 0 and a vertically polarized photon can be the bit 1.

Alice applies her secret rotation $U_A(X)$ on the chosen photon (or photon stream) and then she sends it on to Bob who performs his secret transformation $U_B(X)$ on the photon and sends it back to Alice. Alice performs the inverse transformation ($U^{-1}_A$) and sends it back to Bob who performs ($U^{-1}_B$) thus recovering the message X. The protocol requires that $U_A U_B = U_B U_A$ (i.e.) they are commutative. $U^{-1}_A * U_A = I$, $U^{-1}_B * U_B = I$. Only Alice knows $U_A$ and $U^{-1}_A$, and only Bob knows $U_B$ and $U^{-1}_B$.

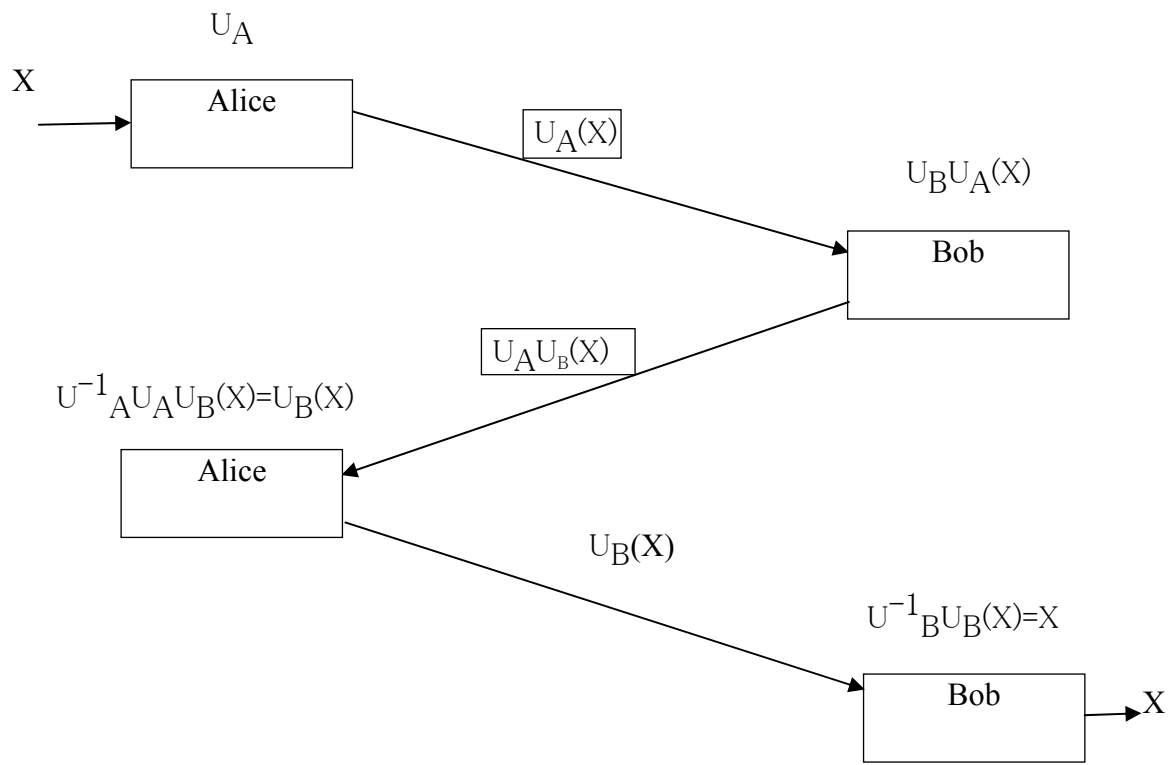

Figure 1. K06 protocol

This is a description of what happens during each stage when we assume that some degree of intensity monitoring is implemented by both Alice and Bob.

*Stage 1.*
Alice has the message X that she wants to send to Bob and does not want anyone to eavesdrop on. So she applies her secret transformation $U_A$ on the message X, thereby creating $U_A(X)$. She then transmits this to Bob. The transmission can be over a secure or non-secure channel.

X= Message
$U_A$= Alice's Transformation
$U_A(X)$= Message with Alice's transformation

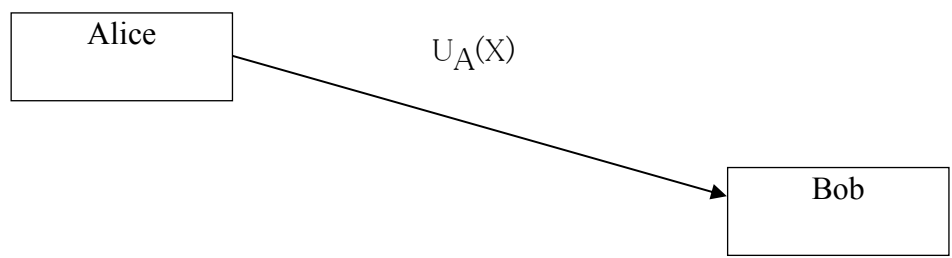

*Stage 2.*
In stage 2 of the protocol Bob now has received Alice's transmission, He takes a reading and there by destroys a certain amount of photons. This step is a precautionary step to check if someone is eavesdropping on their transmissions. Bob then applies his transformation $U_B$ on the message $U_A(X)$ and produces $U_A U_B(X)$ this can also be written as $U_B U_A(X)$ since $U_B$ and $U_A$ are commutative. He then transmits the new message to Alice.

$U_A$= Alice's Transformation
$U_B$= Bob's Transformation
$U_B U_A(X)$= Message with Bob's and Alice's Transformation

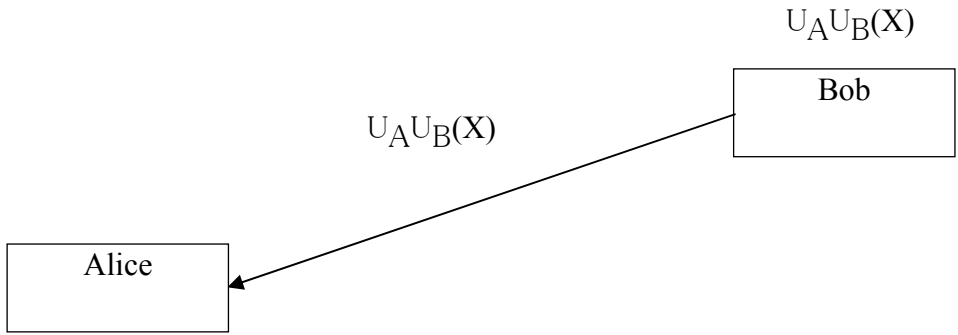

*Stage 3.*
In stage three of the of the protocol, Alice receives the message and takes a power reading, this step is to check if the power level of the transmission is within acceptable levels. If it is within acceptable levels then there was no eavesdropper, otherwise there was an eavesdropper.

After this is done Alice applies her secret inverse transformation $(U^{-1}_A)$ to remove her transformation $U_A$, thereby making the message $U_B(X)$, that is the message now only contains Bob's secret transformation. She then transmits this to Bob who in turn applies his secret inverse transformation on the message and retrieves the original message (X.)

$U^{-1}_A$= Alice's Inverse Transformation
$U^{-1}_B$= Bob's Inverse Transformation
$U_B(X)$= Message with Bob's Transformation

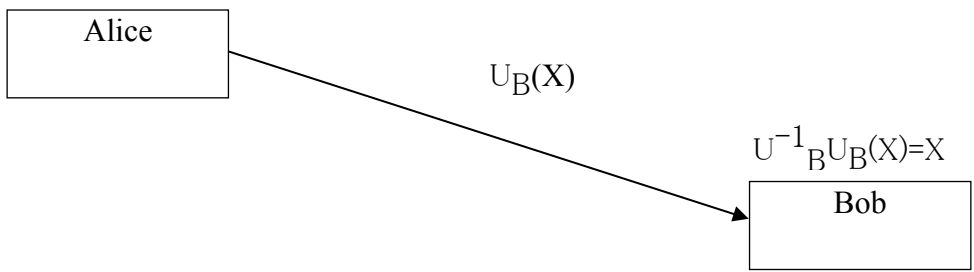

## The Man-in-the-Middle

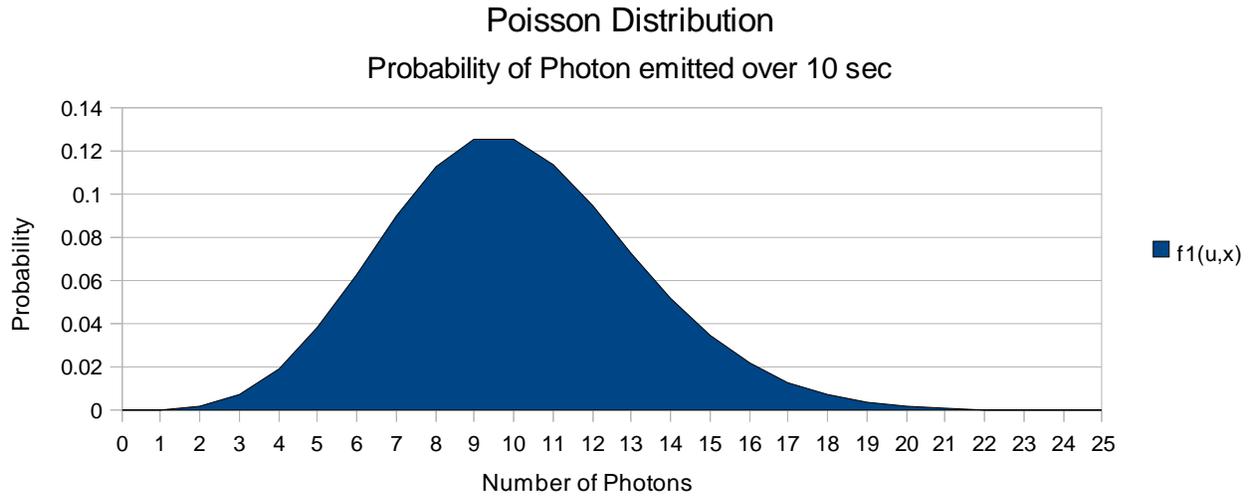

The man-in-the-middle attack can happen in two different ways; the first way is when Eve tries to siphon off photons and use these photons to determine the message, and the second type of attack is when Eve tries to impersonate Alice.

In the K06 protocol, bits that are sent between Alice and Bob are carried by photons but it is not possible to generate one photon [4] at a specific time due to the Heisenberg Uncertainty principle. These photons are generated in burst of waves or pulses. The pulses generated follow a Poisson distribution which is a good model to predict events based on average number of events ($\mu$) known from previous history. The pulse might have only one photon but it is very unlikely.

Suppose Alice is sending a message to Bob using the K06 protocol and Eve wants to listen in on the message, there is a way to detect if Eve is listening on the transmission. Alice and Bob do not know if the line is secure so Alice applies her secret transformation $U_A$ on the pulse (with N photons) and transmits to Bob. Eve in-order to eavesdrop has to siphon off a few number of photons (let's say n is the number of photons that is necessary to measure polarization). The pulse now has N-n number of photons and Bob gets the pulse. The intensity of the light pulse will progressively decrease as it goes across the three stages.

In an intensity-aware implementation of the K06 protocol, let's suppose the two parties use a certain fraction which is equal to N/4 photons. Upon receiving the pulse Bob applies his secret transformation $U_B$ and siphons off a few to measure the intensity and in the process reduces the intensity to N-N/4 photons. Likewise, Alice checks the intensity and reduces it to N/2. Finally, for Bob the beam has intensity of N/4 after he has measured the value at his end. If E does eavesdropping, she would siphon off a large number of photons and both Alice and Bob would get to know that.

When Alice receives the transmission she can check the intensity to determine if someone was eavesdropping on the transmission. Alice can terminate the transmission or can transmit back to Bob if the Alice knows the power of the transmission is too low for Eve to make another reading. This is an example of converting the weakness of the protocol ( i.e. requiring multiple transmissions) to a strength by adding extra security. This three stage transmission is similar multi-located parties and provided an innate protection against man in the middle attack [15].

Eve can try to impersonate Alice and try to send her own message to Bob. However Alice can publish a public hash of her message and Bob only needs to check if the hashes of the message match. When Bob discovers that the hashes do not match, he knows immediately the message has been intercepted by Eve and replaced. The probability of the two different messages to have the same is hash is statistical extremely small.

**Conclusion**

The analysis shows us that the K06 protocol is resilient to man-in-the-middle attack. Should it be known how many photons are needed to determine the polarization, one only needs to transmit at or below that threshold and Eve would not get anything useful by siphoning off the photons.